\documentclass{aa}  
\usepackage{graphicx}
\usepackage{txfonts}
\def\degrees{$^{\circ}$}
\def\srm{$\sigma_{\rm RM}~$}
\def\rmm{$\langle{\rm RM}\rangle~$}

\begin{document}
   \title{The intracluster magnetic field power spectrum in Abell 2382}
   \subtitle{}

 \author{D. Guidetti\inst{1}
         \and
         M. Murgia\inst{1,2}   
         \and 
         F. Govoni\inst{2} 
         \and
         P. Parma\inst{1}
         \and
         L. Gregorini\inst{1,3}
         \and
         H. R. de Ruiter\inst{1,4} 
         \and
         R. A. Cameron\inst{5}
         \and
         R. Fanti\inst{1}
         }
    \offprints{guidet\_s@ira.inaf.it}

   \institute{INAF - Istituto di Radioastronomia,
              Via Gobetti 101, I--40129 Bologna, Italy
              \and
              INAF - Osservatorio Astronomico di Cagliari,
              Loc. Poggio dei Pini, Strada 54, I--09012 Capoterra (CA), Italy
              \and
              Dipartimento di Fisica,
              Univ. Bologna, Via Irnerio 46, I--40126 Bologna, Italy
              \and
              INAF - Osservatorio Astronomico di Bologna,
              Via Ranzani 1, I--40127 Bologna, Italy  
              \and
              Stanford Univ./SLAC, 2575 Sand Hill Road, Menlo Park, CA 94025, USA
              }

   \date{Received; accepted}

  \abstract
  {}
{The goal of this work is to put constraints on the strength and structure of the magnetic field in the cluster of galaxies A2382.
We investigate the relationship between magnetic field and Faraday rotation effects in the cluster, using numerical simulations as a reference for
the observed polarization properties.
   }
 {For this purpose we present Very Large Array observations at 20\,cm and 6\,cm of two polarized radio sources embedded in  A2382, and we obtained 
detailed rotation measure images for both of them.
We simulated random three-dimensional magnetic field models with different power spectra and thus produced synthetic rotation measure images.  
By comparing our simulations with the observed polarization properties of the radio sources,	
we can determine the strength and the power spectrum of intra-cluster magnetic field fluctuations that best reproduce  the observations.
  }
 {The data are consistent with a power law magnetic field power spectrum with the Kolmogorov index $n=11/3$, while the outer 
scale of the magnetic field fluctuations is of the order of 35\,kpc. The average magnetic field strength at the cluster center is about 3\,$\mu$G and 
decreases in the external region as the square root of the electron gas density. The average magnetic field strength in the central 1\,Mpc$^{3}$ is 
about 1\,$\mu$G.}
   {}


   \maketitle
  
%
%________________________________________________________________

\section{Introduction}
\label{sec:intro}

The intra-cluster medium (ICM) in clusters of galaxies is known to possess a magnetic 
field, but its origin and properties are not well known.
The existence of magnetic fields can be demonstrated with different methods of analysis
(see e.g. the review by Govoni \& Feretti 2004, Carilli \& Taylor 2002, 
and references therein).
The strongest evidence for the presence of cluster magnetic fields comes from radio observations. 
Magnetic fields are studied through the synchrotron emission of cluster-wide diffuse sources, 
and from studies of the Faraday rotation  of polarized radio galaxies.
The magnetized plasma that is present between an observer and a radio source changes the properties of the polarized emission 
from the radio source. Therefore, the magnetic field strength can be determined with the help of X-ray observations of the hot gas, 
through the investigation of the Faraday Rotation Measure (RM) of radio sources located inside or behind the cluster.

\begin{figure*}[t]
\centering
\includegraphics[width=18cm]{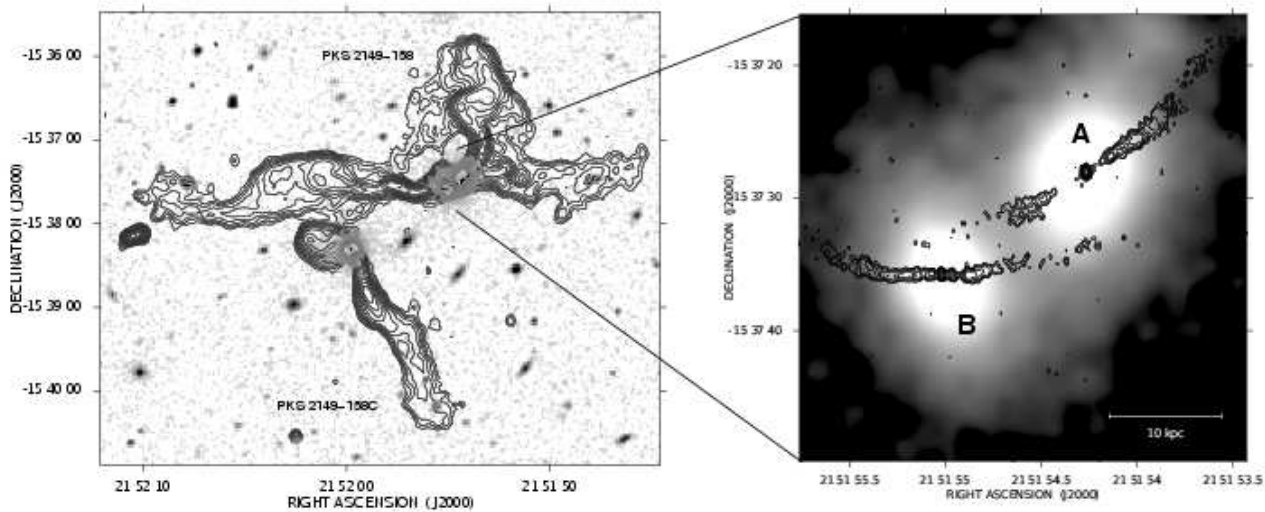}
\caption[]{Left: radio contours of galaxies PKS 2149-158 and PKS 2149-158C obtained at 20 cm 
superimposed on the red DSS2 image. The radio image has been obtained by combining
all the VLA 20\,cm data and by averaging the two IFs within the 20\,cm band. 
The sensitivity (1 $\sigma_{\rm I}$) is 0.052 mJy/beam. The contour levels start at 3$\sigma_{\rm I}$ and are scaled by $\sqrt{2}$; 
the restoring FWHM beam is $5.3''\times5.3''$. Right: zoom of the dumbbell system showing the two radio cores and the respective twin jets. 
The radio contours refer to the A array data at 6\,cm band. Levels start at 0.06 mJy/beam (3$\sigma_{\rm I}$) and increase by $\sqrt{2}$; the angular 
resolution of the radio image is $0.4''\times0.5''$.}
\label{20cm}
\end{figure*}
RM studies of radio galaxies in clusters have been carried out using either statistical samples (e.g. Lawler \& Dennison 1982, Vall\'ee et al. 1986,
Kim et al. 1990, Kim et al. 1991, Clarke et al. 2001)  or individual clusters. In the latter case one analyses detailed high resolution 
RM images (e.g. Perley \& Taylor 1991, Taylor \& Perley 1993, Feretti et al. 1995, Feretti et al. 1999, Govoni et al. 2001, Taylor et al. 2001, 
Eilek \& Owen 2002, Pollack et al. 2005, Govoni et al. 2006).
These data are usually consistent with central magnetic field strengths of a few $\mu$G, but stronger fields are found in the inner regions 
of relaxed cooling core clusters, and can reach values of 10~--~40 $\mu$G (see e.g. Taylor et al. 2002). 
Both for interacting and relaxed clusters the RM distribution of radio galaxies is generally patchy, indicating that cluster magnetic
fields have structures on scales as small as 10 kpc or less.\\
On the basis of the available RM images, increasing attention is given in the literature to the power spectrum of the intra-cluster magnetic field fluctuations.
Several studies (En{\ss}lin \& Vogt 2003, Murgia et al. 2004) have shown that detailed
RM images of radio galaxies can be used to infer not only the cluster magnetic field strength, but also the cluster magnetic field power spectrum. 
The analysis of Vogt \& En{\ss}lin (2003, 2005) suggests that the power spectrum is of the Kolmogorov type if the auto-correlation length of the magnetic 
field fluctuations is of the order of a few kpc. However, Murgia et al. (2004) pointed out that shallower magnetic field power spectra are possible if the 
 magnetic field fluctuations extend out to several tens of kpc. Recently, Govoni et al. (2006), who used the numerical approach developed by Murgia 
et al. (2004), derived the power spectrum of the intra-cluster magnetic field fluctuations in A2255, and found that  the field strength declines from the cluster 
center outwards, with an average field strength of about 1.2 $\mu$G over the central 1 Mpc$^3$. They also showed that in order to explain both the 
observed RM of the radio galaxies in A2255 and polarization levels of the radio halo present in this cluster (Govoni et al. 2005), the maximum scale of the 
magnetic field fluctuations must be of the order of hundreds of kpc with a steepening of the power spectrum from the cluster center to the periphery.

In this paper we present Very Large Array (VLA\footnote{The Very Large Array is a facility of the National Science Foundation, operated under 
cooperative agreement by Associated Universities, Inc.}) observations at 20\,cm and 6\,cm of the three polarized radio galaxies
PKS 2149-158 (A and B) and PKS 2149-158C in the cluster Abell 2382; the first two (A and B) form a dumb-bell system.
A2382 is an ideal  case to study  RM along different lines-of-sight  because PKS 2149-158 and PKS 2149-158C are extended and highly polarized radio 
galaxies, located respectively at 4.8$'$ and 4.3$'$ from the cluster center.
Because the radio sources under investigation are quite extended, both in angular and linear size, they are ideal targets for an 
analysis of the rotation measure distribution: detailed RM images can be constructed which can serve as the basis of an accurate study of magnetic 
field power spectrum. 

We follow the numerical approach proposed by Murgia et al. (2004), i.e. we study the polarization properties of the radio galaxies, at the same time
making use of the cluster X-ray information.

The paper is organized as follows.
In Sect.~\ref{sec:obs} we discuss details about the radio observations and the data reduction.
In Sect.~\ref{sec:IP} we present the total intensity and polarization properties of the radio galaxies at 20 and 6 cm. We also describe the morphology of the 
sources and the discrete features found in the total intensity images.
In Sect.~\ref{sec:X} we present the X-ray environment in which the radio galaxies are embedded.
In Sect.~\ref{sec:RM} we show  the RM images, discuss the results, and discuss the cluster magnetic field.
In  Sect.~\ref{sec:RMan} by following the same approach as in Murgia et al. (2004), we introduce the multi-scale magnetic field modelling used to determine
the intra-cluster magnetic field strength and structure, and we show the results obtained with different
 configurations of the  magnetic field power spectrum slope.
Finally, Sect.~\ref{sec:sum} summarizes our main conclusions.

Throughout this paper we assume a $\Lambda$CDM cosmology with
$H_0$ = 71 km s$^{-1}$Mpc$^{-1}$,
$\Omega_m$ = 0.3, and $\Omega_{\Lambda}$ = 0.7.
At the distance of A2382 ($z=0.0618$, Struble \& Rood 1999), 1 arcsec corresponds to 1.17 kpc.

%__________________________________________________________________

\begin{figure*}
\centering
\includegraphics[angle=-90, width=15cm]{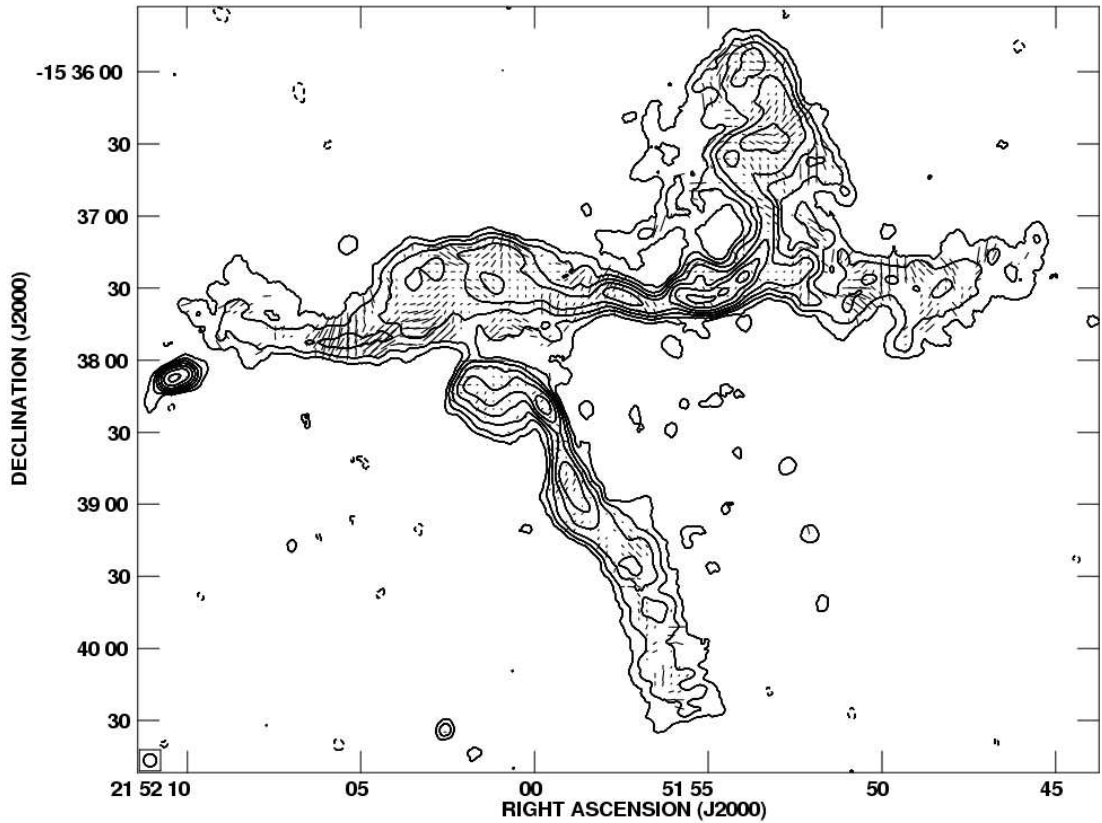}
\caption[]{Source PKS 2149-158 (A and B) and PKS 2149-158C:
total intensity contours and polarization vectors at 1.46 GHz. 
The angular resolution is $5.3'' \times 5.3''$.
The first contour level is drawn at -3$\sigma_{\rm I}$ and the other contour levels start at 3$\sigma_{\rm I}$ and are spaced by a factor of 2.	
The lines give the orientation of the
electric vector position angle (E-field) and are proportional 
in length to the fractional polarization ($10'' \simeq50$\%).}
\label{2149}
\end{figure*}

\begin{figure*}
\centering
\includegraphics[angle=-90, width=15cm]{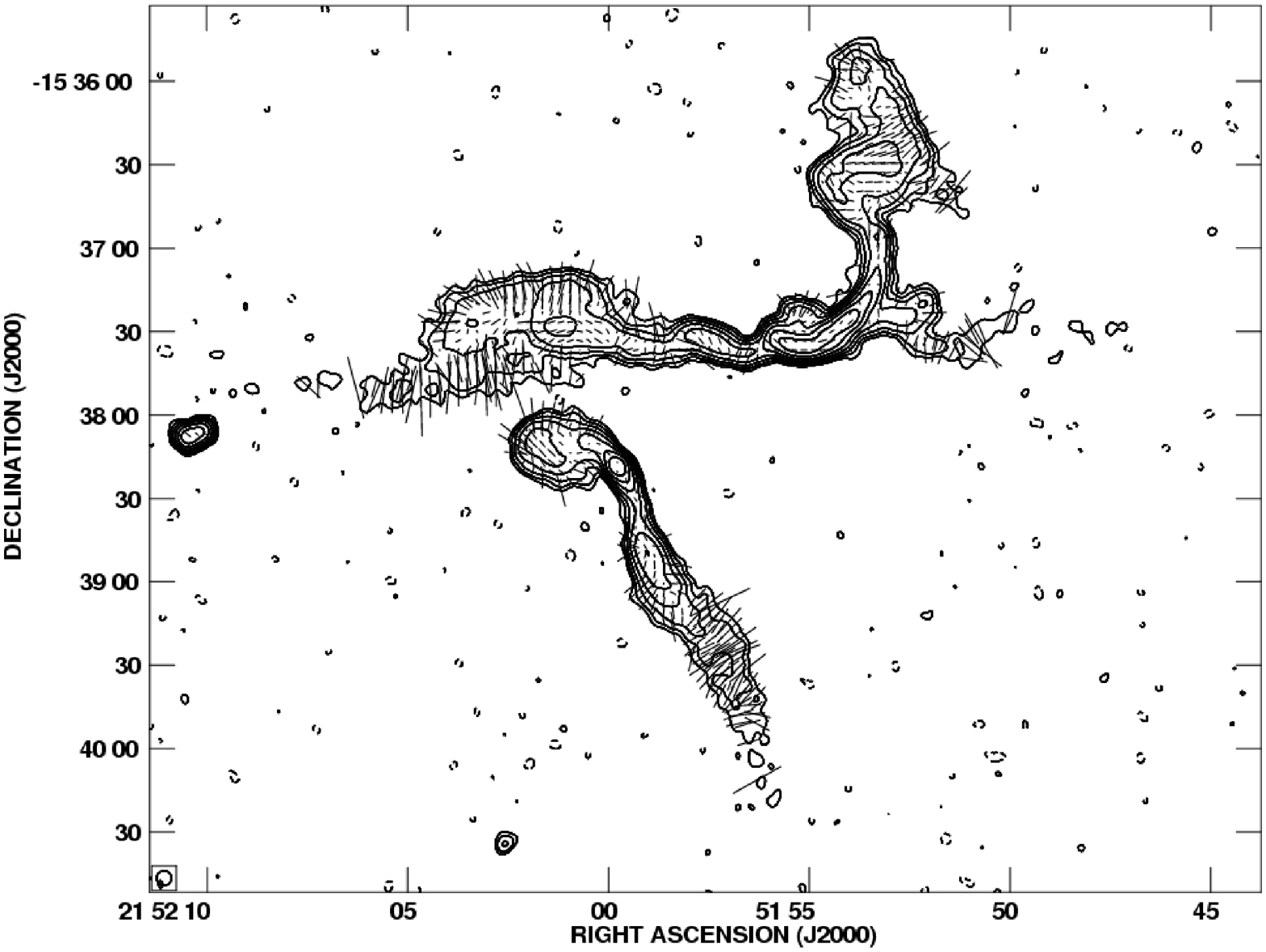}
\caption[]{Source PKS 2149-158 (A and B) and PKS 2149-158C:
total intensity contours and polarization vectors at 4.88 GHz. 
The angular resolution is $5.3'' \times 5.3''$.
The first contour level is drawn at -3$\sigma_{\rm I}$ and the other contour levels start at 3$\sigma_{\rm I}$ and are spaced by a factor of 2.	
The lines give the orientation of the
electric vector position angle (E-field) and are proportional 
in length to the fractional polarization ($10'' \simeq50$\%).}
\label{2149c}
\end{figure*}

\section{Radio observations and data reduction}
\label{sec:obs}
The radio sources PKS 2149-158 (A and B) and PKS 2149-158C were observed at the 20 and 6\,cm bands, in all VLA configurations.
The details of the observations are provided in Tab.\,1.
The observations in the different arrays were made between November 1986 and December 1987.

\begin{table} 
\label{data}
\caption{Summary of the VLA observations.}          
\centering          
\begin{tabular}{ c c c c }     % 4 columns 
\hline\hline       
 $\nu$      & Bandwidth & Config.  & Duration \\
  (GHz)     & (MHz)     &          & (Hours)  \\
\hline                    
1.45/1.65    & 25 & A & 6.6 \\ 
1.46/1.66    & 25 & B &  6.3 \\
1.46/1.66    & 50 & C &  3.6 \\
1.46/1.66    & 50 & D &  0.9 \\
4.82/4.87    & 25 & A & 6.1  \\
4.82/4.87    & 25 & B & 5.8  \\
4.83/4.88    & 50 & C & 14.6 \\
4.83/4.88    & 50 & D & 11.3  \\
\hline
\multicolumn{4}{l}{The pointing position (J2000) is at RA=21$^{h}$51$^{m}$57\arcsec} \\
\multicolumn{4}{l}{and DEC=-15\degr37\arcmin23\arcsec.}\\
\end{tabular}
\end{table}

The flux densities were brought on the scale of Baars et al. (1990) using 3C\,286 as primary flux density calibrator. 
The same calibrator has been used as the absolute reference for the electric vector polarization angle.
The phase calibrators were nearby point sources observed at intervals of about 30 minutes.

Calibration and imaging were performed with the NRAO Astronomical Image Processing System (AIPS), 
following standard procedures. Several cycles of self-calibration and CLEAN were applied to remove residual phase variations.
Instrumental calibration of the polarization leakage terms was obtained using the phase calibrators, which were observed over 
a wide range in parallactic angle. The calibration of the absolute polarization angle was obtained by assuming
for the source 3C\,286 a R-L phase difference of 66\degr~at both 20 and 6\,cm. 

The ($u,v$) data at the same frequencies but from different configurations were first handled separately and then combined 
 to improve uv-coverage and sensitivity. The separate data sets were individually flagged, flux calibrated, and polarization
 calibrated before combination.  We combined all arrays at both 20 cm and  6 cm. Each combined data set
 was then self-calibrated.

Images of polarized intensity $P=(Q^2+U^2)^{1/2}$, 
fractional polarization $FPOL=P/I$ and position angle of polarization
$\Psi=0.5\tan^{-1}(U/Q)$ were derived from the $I$, $Q$, and $U$ images.

\begin{table}[t]
\label{thesource}
\caption[]{Basic properties of PKS 2149-158 and PKS 2149-158C.}
\begin{tabular}{ll}
\hline\hline
\noalign{\smallskip}
 PKS 2149-158: &\\
Position (J2000) \hfill  radio core A   & 21$^{h}$51$^{m}$54\farcs3  ~~-15\degr37\arcmin28\farcs0  \\
~~~~~~~~~~~$''$      \hfill radio core B& 21$^{h}$51$^{m}$55\farcs0  ~~-15\degr37\arcmin35\farcs8  \\

Redshift                                & 0.062\\
Total flux density at 20\,cm               & 424 mJy\\
Total radio luminosity at 20\,cm    & $10^{24.6}$ W/Hz\\
Overall spectral index $\alpha^{6\rm\,cm}_{20\rm\,cm}$ & 0.95\\
Radio source largest linear size        & 410 kpc\\
\noalign{\medskip}
\hline

PKS 2149-158C: &\\
Position (J2000) \hfill radio core      & 21$^{h}$51$^{m}$59\farcs8  ~~-15\degr38\arcmin18\farcs2  \\
Redshift                                & 0.060\\
Total flux density at 20\,cm            & 138 mJy\\
Total radio luminosity at 20\,cm        & $10^{24.0}$ W/Hz\\
Overall spectral index $\alpha^{6\rm\,cm}_{20\rm\,cm}$  & 0.86\\
Radio source largest linear size        & 210 kpc\\
\noalign{\smallskip}
\hline

\multicolumn{2}{l}{\scriptsize We use the convention $S_{\nu}\propto
\nu^{-\alpha}$}\\
\end{tabular}
\label{source}
\end{table}

\section{Total intensity and polarization properties}
\label{sec:IP}
Radio images at 20\,cm and 6\,cm were obtained by combining all the VLA arrays and by averaging the two IFs.
The left panel of Fig.\,\ref{20cm} shows the total intensity contours of the 20\,cm image, which was restored with a FWHM beam of $5\farcs 3$,  
overlayed onto the optical image\footnote{htpp://archive.eso.org/dss/dss} from the red Palomar Digitized Sky Survey 2. A zoom over the
 dumb-bell system is presented on the right  panel of Fig.\,\ref{20cm}, where we show the total intensity contours of the 6\,cm image, 
which was restored with a FWHM beam of $0\farcs 4 \times 0\farcs 5$.

The basic properties of PKS2149-158 and PKS2149-158C are given in Tab.~\ref{source}. Flux densities were estimated after having applied 
the primary beam correction, by integrating in the same area the surface brightness down to the noise level.

\begin{table*}
\caption{Total and polarization intensity radio images, for each individual IF, restored with a FWHM beam of $5.3'' \times 5.3''$.}
\label{mappe}
\centering
\begin{tabular} {c c c c c c c c c c} 
\hline\hline
Config.    & $\nu$    &Beam      & $\sigma_{\rm I}$  & $\sigma_{\rm Q}$ & $\sigma_{\rm U}$  \\
             & (GHz)         &($''$)    & (mJy/beam)   & (mJy/beam)  & (mJy/beam)   \\ 
\hline
A+B+C+D &  1.46 & 5.3$\times$5.3 & 0.054 & 0.024 & 0.024 \\
$''$ &  1.66 & $''$ & 0.047 & 0.027 & 0.027  \\
$''$ & 4.83 & $''$ & 0.022 & 0.019 & 0.019  \\
$''$ & 4.88 & $''$ & 0.020 & 0.019 & 0.019 \\ 
\hline
%\multicolumn{9}{l}{Col. 1: Array; Col. 2: Observation frequency; Col. 3: FWHM beam; Col. 4, 5, 6: RMS noise in individual IF images }\\
%\multicolumn{9}{l}{of the I, Q, U images.} \\
\end{tabular}
\end{table*}
For the purpose of the polarization and RM analysis, intensity and polarization images were produced also for each IF separately.
The relevant information on these images is listed in Tab.~\ref{mappe}. 
Total intensity contours and polarization vectors at 1.46\,GHz and 4.88\,GHz
are shown in Figs.\,\ref{2149} and \ref{2149c} respectively.
Vectors represent  the orientation of the projected 
E-field and are proportional in length to the fractional polarization.
In the fractional polarization images \textit{FPOL}, we included only the points with  $I>5\sigma_{\rm I}$.

In the following we give a brief description of the individual sources.

\subsection{PKS 2149-158 (A and B)}
\label{subs:PKS}
PKS 2149-158 (FR class I) is a double system composed of two nearly equally bright elliptical galaxies in a common envelope 
("dumb-bell'' galaxy).
The radio source was first mapped by Parma et al. (1991) at 1.4 GHz. Both galaxies of the dumb-bell system are radio-emitters, forming 
a double twin jet system like 3C\,75 (Owen et al. 1985). The two radio cores, labelled A and B in  Fig.\,\ref{20cm}, are separated in projection by
 $13\farcs5$ which corresponds to 15.8\,kpc. Their position is reported in Tab.~\ref{source}.

The radio morphology of PKS 2149-158, which shows regular large amplitude oscillations,  
is rather unusual and can be interpreted in terms of two distinct radio sources whose jets are strongly interacting. The true (three-dimensional) source structure
is undoubtedly even more complex, because of the radio jets may very well bend along the line-of-sight.
The maximum angular size of the radio source at 20\,cm is about  $5.8\arcmin$ ($\simeq$410\,kpc).

The magnetic field configuration, as traced by the 6\,cm images, is initially transverse in the jets and becomes circumferential in the lobes. 
However, there are systematic differences between the linear polarization at 20\,cm and 6\,cm because of the Faraday rotation effect. 
At the higher frequency (Fig.\,\ref{2149c}), the $\vec{E}$-vectors in both jets and lobes are rather ordered and the degree of polarization is high.
In contrast, the $\vec{E}$-vectors at 20\,cm  (Fig.\,\ref{2149}) in the center of the lobes, have a much more chaotic distribution.
The mean fractional polarization is $\simeq$12$\%$ at 20\,cm and $\simeq$23$\%$ at 6\,cm. The signal is affected by beam depolarization at longer wavelength.

\subsection{PKS 2149-158C}
\label{subs:PKSC}
The radio source (FR class I) is associated with a single elliptical galaxy and is unrelated to the dumb-bell system.
Its total extent at 20\,cm is about $3\arcmin$ ($\simeq$ 210\,kpc).
The source has a double asymmetric  structure with the two jets emanating from the core to the north-east and south-west. The north jet and 
south jet extend out to $9.2\arcsec$ and $18\arcsec$ respectively from the core. Of course, the north-east jet may appear
distorted by projection effects and probably bends along the line-of-sight.

Figs.\,\ref{2149} and \ref{2149c} show the polarization images of the source at 20\,cm and  6\,cm. 
The magnetic field configuration, as traced by the 6\,cm images, is initially transverse to the jets, parallel to the southern lobe, and 
circumferential in the northern lobe. The mean fractional polarization is $\simeq$ $7\%$ at 20 \,cm and $\simeq$ 25$\%$ at 6\,cm. As for the dumb-bell system, the signal is affected by beam depolarization at longer wavelength.

\section{X-ray environment}
\label{sec:X}
\begin{figure*}
\centering
\includegraphics[height=8.5cm]{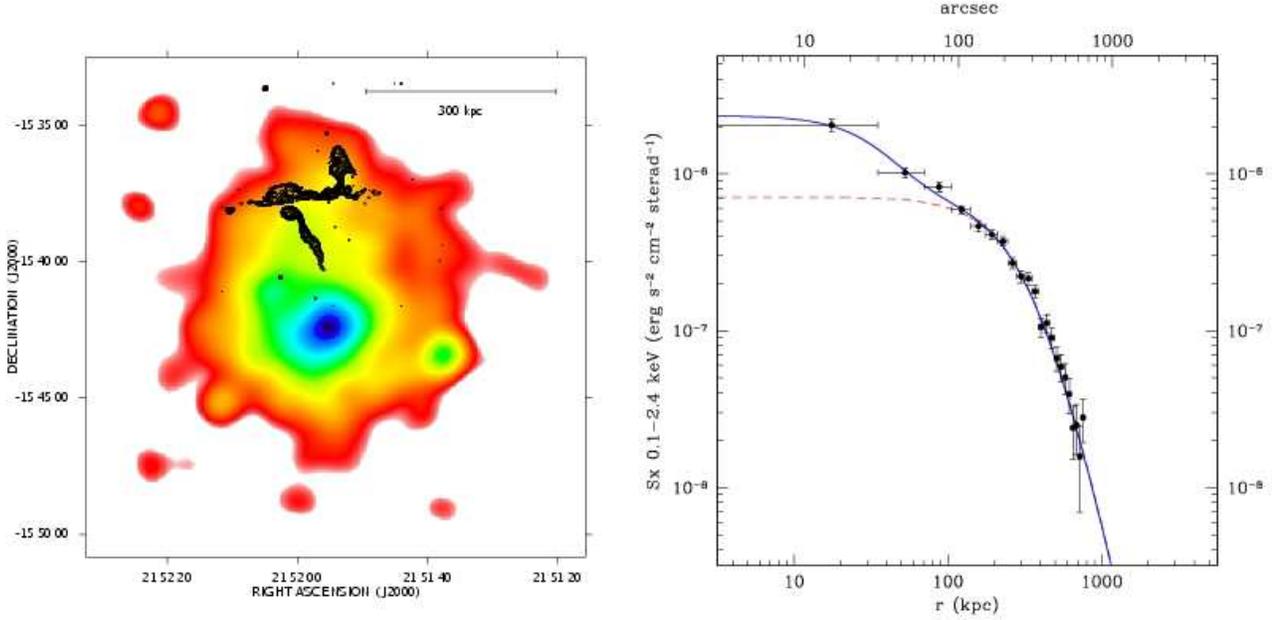}
\caption[]{Left: radio contours of galaxies PKS 2149-158 and PKS 2149-158C  at 6 cm 
superposed on the Rosat PSPC X-ray image. The radio image has been obtained by combining
all the VLA arrays and by averaging the two IFs of the 6\,cm band. 
The sensitivity (1 $\sigma_{\rm I}$) is 0.015 mJy/beam.
The contour levels start at 3$\sigma_{\rm I}$ and are scaled by $\sqrt{2}$;  the restoring FWHM beam is $5\farcs 3\times 5\farcs 3'$.
Right: surface brightness profile  of A2383 in the band 0.1-2.4\,keV band. The dashed and the solid  lines represent the best fit of, respectively,
the single $\beta$-model and the double $\beta$-model described in the text.}
\label{rosat}
\end{figure*}

The cluster A2382 has been observed in  X-rays with the Rosat  satellite.
The left panel of Fig.\,\ref{rosat} shows total intensity contours at 6\,cm 
superposed on the  Rosat  PSPC archive  image (800227p) of  A2382. 
The X-ray image represents intensity in the 0.1-2.4\,keV band. It has been corrected for the background, 
divided by the exposure map (a $\simeq17$\,ksec exposure) and smoothed with a Gaussian of $\sigma =30\arcsec$.
The centroid of the image is located at RA=21$^{h}$51$^{m}$55\arcsec  DEC=-15\degr42\arcmin26\arcsec; the  X-ray emission  
extends up to more than $15\arcmin$.  
The radio-X  overlay shows that the two radio sources 
PKS 2149-158 and PKS 2149-158C are offset to the north of the cluster centre by about $4.8\arcmin$ (340\,kpc)
and 4.3$\arcmin$ (300\,kpc) respectively.

\subsection{X-ray surface brightness profile} \label{subs:fitsx}
The right panel of Fig.\,\ref{rosat} shows the  X-ray surface brightness ($S_{X}$) profile of A2382.
The profile was obtained by averaging the 0.1-2.4\,keV Rosat image (corrected for the background and divided by the exposure) in concentric
annuli of $30\arcsec$ in size, centered on the X-ray centroid.
Point sources have been masked. 
We converted the X-ray surface brightness from  counts/skypixel/sec to erg cm$^{-2}$sec$^{-1}$sterad$^{-1}$ by using the PIMMS\footnote{http://heasarc.gsfc.nasa.gov/docs/software/tools/pimms.html} software (Mukai 2007).
In this conversion the X-ray emission was approximated by a Raymond-Smith model with a mean cluster temperature kT$\simeq$2.9\,keV 
(Ebeling et al. 1996), metal solar abundance Z=0.6 and photoelectric absorption column density nH=4.1$\times10^{20}$cm$^{-2}$.

Because a strong emission core is present in the inner 100\,kpc of the cluster,
the observed  $S_{X}$ cannot be described by the simple $\beta$-model (Cavaliere \& Fusco-Fermiano, 1976) for the gas density:
\begin{equation} \label{beta}
n_e(r) = n_0 (1+r^2/r_c^2)^{-\frac{3}{2}\beta},
\end{equation}
where $r$, $n_0$ and $r_c$  are the 
distance from the cluster X-ray centroid, the central electron density, 
and the cluster core radius, respectively.

Therefore, we first fitted the radial X-ray surface brightness only for $r>100$\,kpc using the $\beta$-model with the three  free parameters 
$\beta_{\rm EXT}$, $r_{c_{\rm EXT}}$, $n_{0_{\rm EXT}}$, resulting in a best fit with  
$\chi_{\rm EXT}^2=11.8$, for 16 degrees of freedom.
We then used these values in a subsequent fit of the $S_{X}$ profile by combining a double 
 $\beta$-model:
\begin{equation} \label{beta2}
n_e(r) = n_{0_{\rm INT}}(1+r^2/r_{c_{\rm INT}}^2)^{-\frac{3}{2}\beta_{\rm INT}} + n_{0_{\rm EXT}}(1+r^2/r_{c_{\rm EXT}}^2)^{-\frac{3}{2}\beta_{\rm EXT}}
\end{equation}
The double $\beta$-model is based on 6 parameters ($\beta_{\rm INT}$, $r_{c_{\rm INT}}$, $n_{0_{\rm INT}}$, $\beta_{\rm EXT}$, $r_{c_{\rm EXT}}$, $n_{0_{\rm EXT}}$ ) 
of which the last three are fixed at the values calculated by the previous fit. In this case, the final $\chi_{\rm INT}^2$ is 15.0 for 19 degrees of freedom.
The fits of the two $\beta$-model (single and double) are shown respectively as dashed and solid line in Fig.\,\ref{rosat} (right panel). The best fit 
parameters are listed in Tab.~\ref{xfit}.
Overall the model fit is very good. It gives a central gas density of $n_0\equiv n_{0_{\rm INT}}+ n_{0_{\rm EXT}}=5\times10^{-3}$cm$^{-3}$ and a outer core radius of 373 \,kpc  with 
a outer $\beta_{\rm EXT}$ of 0.9.
The inner  $\beta_{\rm INT}>0.7$ and $r_{c_{\rm INT}}>11$\,kpc are lower limits. This means that larger and larger values of the 
 inner core radius and $\beta$ still satisfy the data, provided that these two parameters grow together.

\begin{table}[b]
\caption{A2382 X-ray surface brightness profile best fitting parameters.}
\label{xfit}
\centering
\begin{tabular} {c c c c} 
\hline\hline
Parameter  & Value & 1-$\sigma$ range & Units  \\
\hline
$n_{0_{\rm EXT}}$ & 1.2 & 1.0$-$1.5  & 10$^{-3}$cm$^{-3}$ \\ 
$r_{c_{\rm EXT}}$ & 373 & 244$-$625 & kpc \\
$\beta_{\rm EXT}$ & 0.9 & 0.7$-$1.6 &  \\
$\chi_{\rm EXT}^2$/NDF & 0.74 (11.8/16) & & \\
\hline
$n_{0_{\rm INT}}$ & 3.8 & 1.8$-$ 20 &  10$^{-3}$cm$^{-3}$ \\ 
$r_{c_{\rm INT}}$ & 65.7 & $>11$ &  kpc \\ 
$\beta_{\rm INT}$ & 1.7 & $>0.7$ & \\ 
$\chi_{\rm INT}^2$/NDF & 0.79 (15.0/19) & & \\
\hline
\end{tabular}
\end{table}

\section{Rotation measure images} \label{sec:RM}

Polarized radiation from cluster and background radio galaxies 
may be rotated by the Faraday effect if magnetic fields are present in the intra-cluster medium.
Linearly polarized electromagnetic radiation passing through a magnetized ionized medium suffers a rotation
of the plane of polarization:

\begin{equation}
 \Psi_{Obs}(\lambda) = \Psi_{int} + (\lambda)^2 \times RM 
\label{rm}
\end{equation}
where  $\Psi_{Obs}(\lambda)$  is the position angle observed at a wavelength  $\lambda$,
$\Psi_{int}$ is the intrinsic position angle and  the Rotation Measure (RM) is related to the electron 
density ($n_e$), the magnetic field along the line-of-sight ($B_{\parallel}$),
and the path-length (L)
through the intracluster medium according to:
\begin{equation}
RM_{\rm~[rad/m^2]}=812\int_{0}^{L_{[kpc]}}n_{e~[cm^{-3}]}B_{\parallel~[\mu G]}dl
\label{equaz}
\end{equation}

The position angle of the plane of polarization is an observable quantity,
therefore, images of rotation measure can be obtained by a linear fit of the polarization
angle as a function of $\lambda^2$ (see e.g. AIPS task RM or the PACERMAN algorithm by Dolag et al. 2005).
As is well known, determination of the rotation measure is 
complicated because of n$\pi$ ambiguities in the observed $\Psi_{Obs}$.
Removal of these ambiguities requires observations at many frequencies that are well-spaced
 in $\lambda^{2}$.
\begin{figure*}
\centering
\includegraphics[width=18cm]{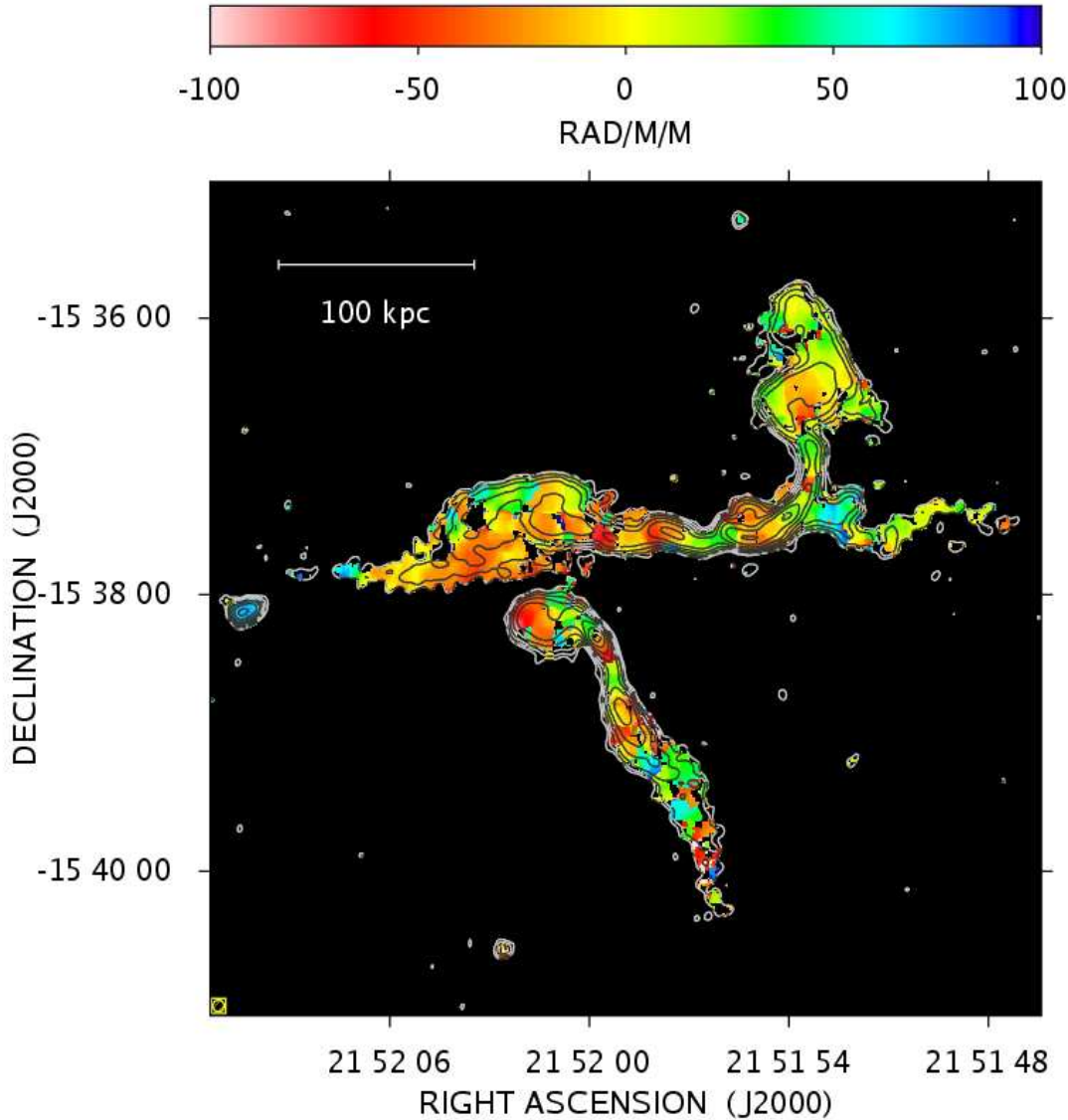}
\caption[]{Images of the rotation measure  
computed using the polarization Q and U  maps at
the frequencies 1.46, 1.66, 4.83 and 4.88 GHz with a resolution of $5\farcs 3 \times 5\farcs 3$.
Contours refer to the total intensity image at 6\,cm. The sensitivity (1 $\sigma_{\rm I}$) is 0.015 mJy/beam;
the contour levels start at $3\sigma_{\rm I}$ and are scaled by a factor of 2.} 
\label{rmfig}
\end{figure*}

\begin{figure*}[t]
\centering
\includegraphics[width=18cm]{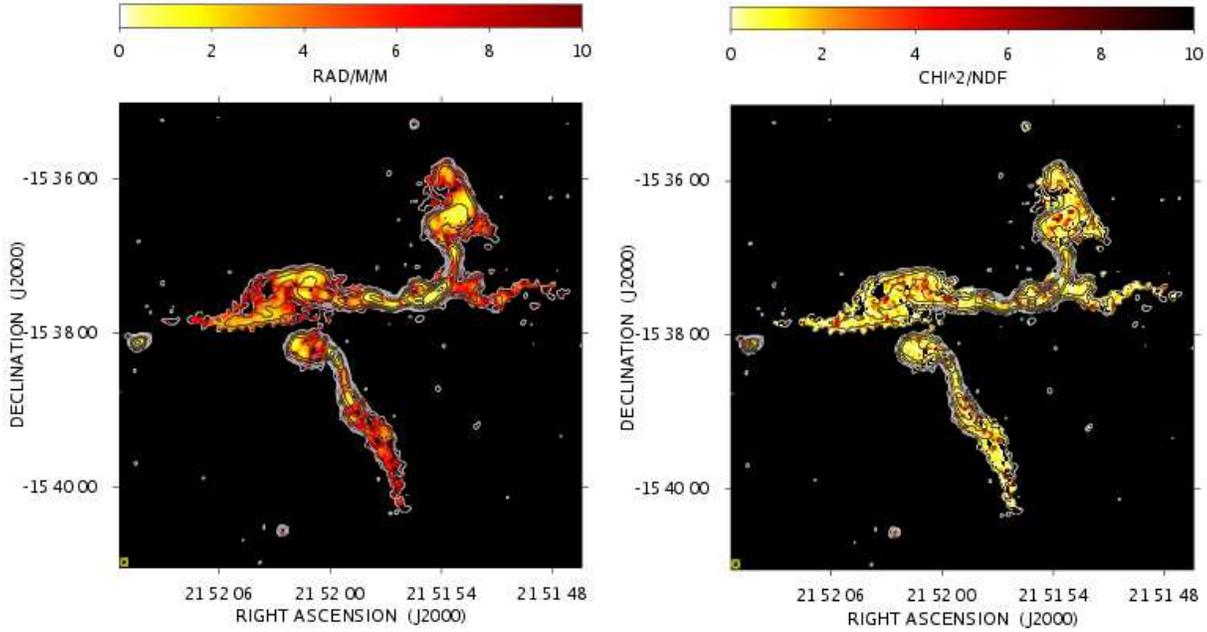}
\caption[]{Left: RM error image. Right:  $\chi^2$-reduced map of the fit computed to obtain RM image. 
Contours refer to the total intensity image at 6\,cm. The sensitivity (1 $\sigma_{\rm I}$) is 0.015 mJy/beam;
the contour levels start at $3\sigma_{\rm I}$ and are scaled by a factor of 2.} 
\label{noise}
\end{figure*}

\begin{figure*}
\centering
\includegraphics[width=16cm]{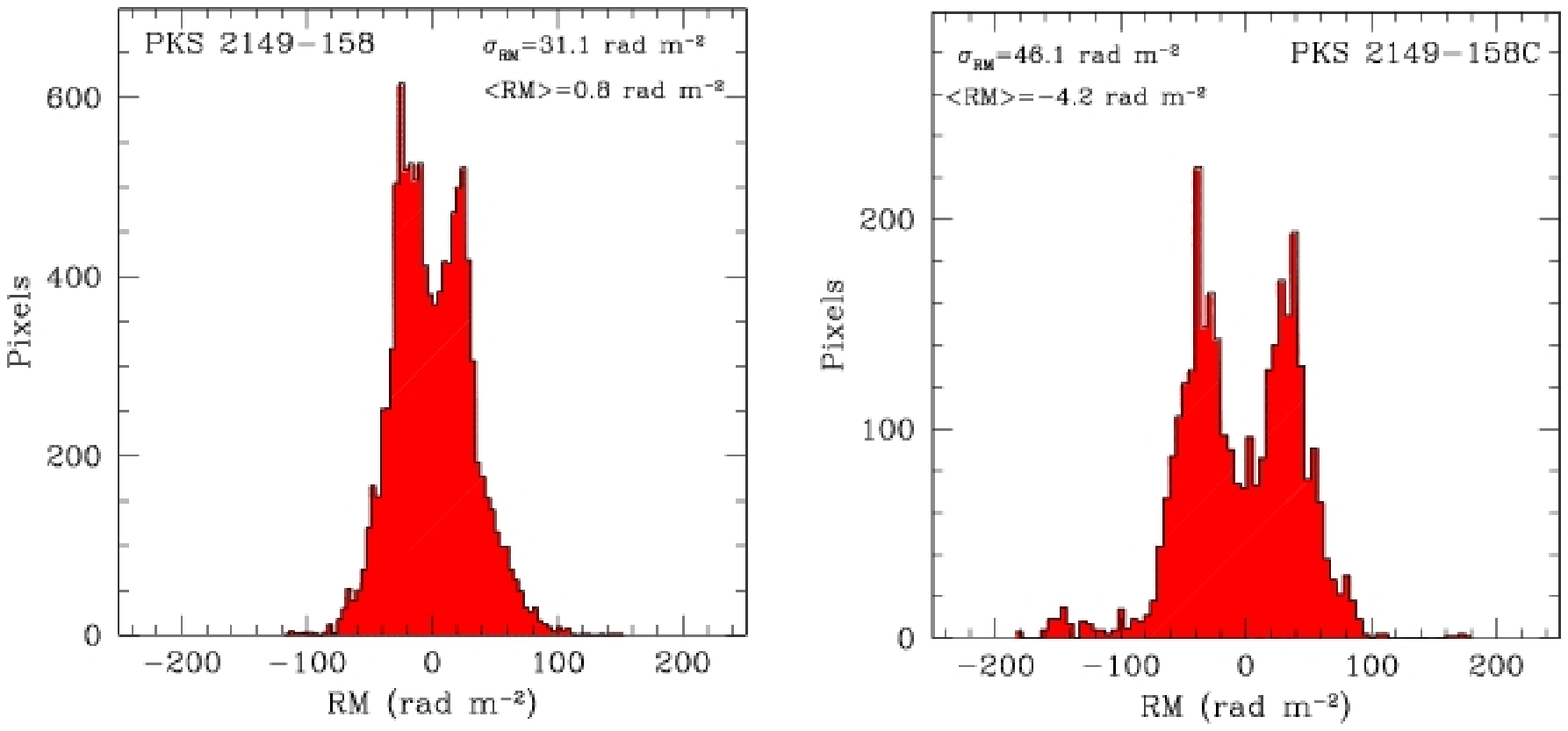}
\caption[]{Histograms of the rotation measure images of  PKS 2149-158 (left) and  PKS 2148-158C (right).} 
\label{histo}
\end{figure*}

We implemented an RM fit algorithm in the FARADAY tool (Murgia et al. 2004).
Given the U and Q maps at each frequency as inputs the task \textit{UQ\_to\_RM} 
produces the RM and the intrinsic polarization angle, both with relative error maps,
and a $\chi^2$ map, obtained by fitting the observed polarization angle images. 
In order to improve the RM image, the algorithm can be iterated several times, by using the RM information 
in the high signal-to-noise region and thus improving computations in adjacent bad pixels.

Fig.\,\ref{rmfig} shows the image of the rotation measure of PKS 2149-158 and PKS 2149-158C
 computed using the polarization Q and U maps at the
frequencies 1.46, 1.66, 4.83 and 4.88\,GHz, with a resolution of $5\farcs3$. Contours refer to the 
total intensity image at 6\,cm.
The RM was calculated only in pixels with I$>$3$\sigma_{\rm I}$ at 6 cm.
In the RM image thus obtained we blanked all pixels with an fitting error 
greater than 10\,rad/m$^2$.
The bulk of RM values range from about $-$100 rad/m$^2$ up to 100 rad/m$^2$. 
Alternating positive and negative RM patches are apparent in each source with
RM fluctuations down to scales of a few kpc.

In order to quantify the goodness of the fit of the rotation measure image, we show, in Fig.\,\ref{noise}, the RM error image
and the $\chi^2$ map. The average error of the RM as given by the fit procedure is about 5 rad/m$^2$ while the
average reduced $\chi^{2}$ is 1.4.

We can characterize the RM distribution in terms of a mean (\rmm) and root mean square (\srm).
Fig.\,\ref{histo} shows the 
histograms of the RM distribution for the two sources.
The distributions of RM have approximately zero mean; in detail:
for PKS 2149-158 we found \rmm=0.8\,rad/m$^2$, and  \srm=31\,rad/m$^2$,
while  for  PKS 2149-158C \rmm=$-$4.2\,rad/m$^2$, and \srm=46\,rad/m$^2$.

\begin{figure*}[t]
\centering
\includegraphics[width=18cm]{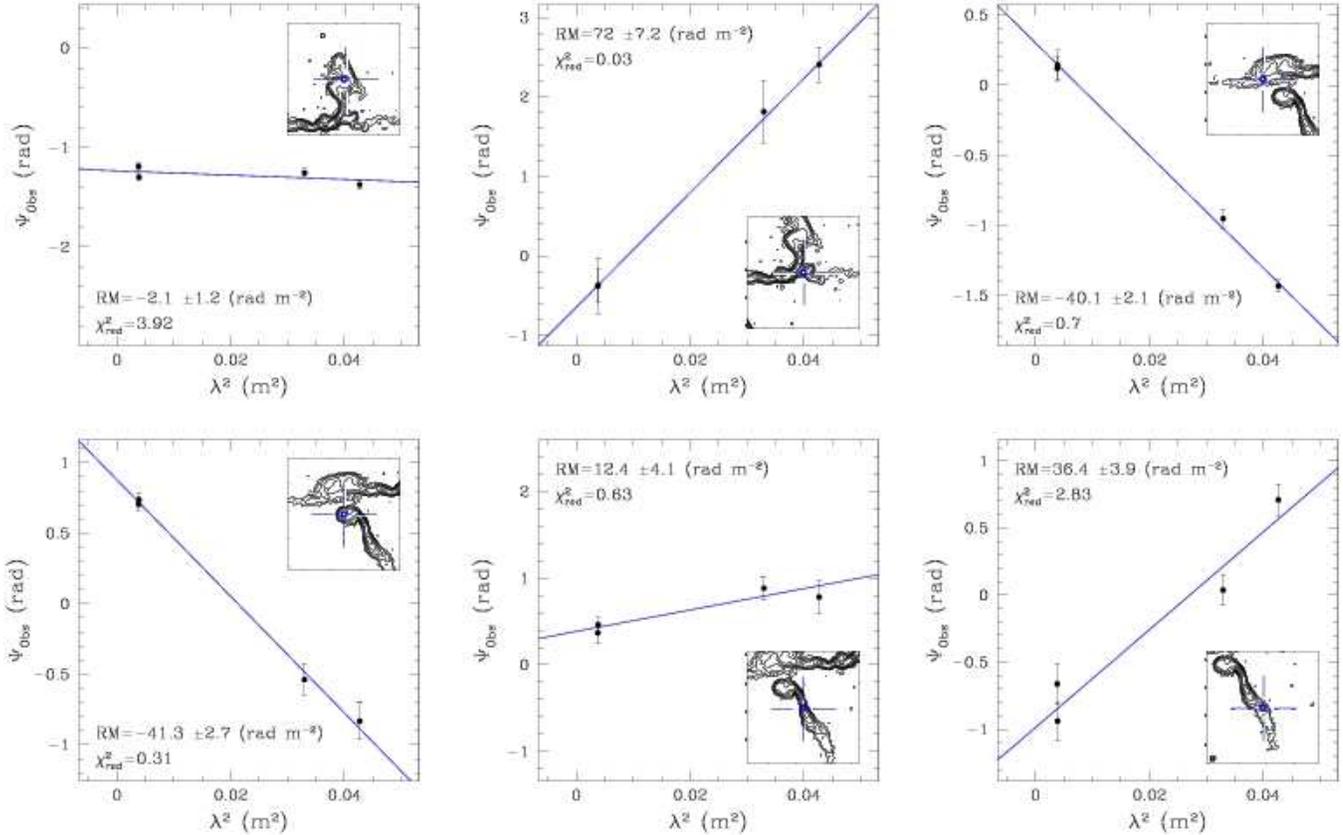}
\caption[]{Sample plots of the $\vec{E}$-vector position angle $\Psi_{obs}$ against $\lambda^2$ at different locations in PKS 2149-158 (top panels) 
and PKS 2148-158C (bottom panels). The exact position of the points in the sources is shown in the insets of the individual panels. 
The solid line represent the best fit of the 
$\lambda^{2}$-law to the data.} 
\label{rmfit}
\end{figure*}

In order to  verify the polarization angle linearity with $\lambda^2$,
in the RM map we selected some pixels corresponding to source
regions with high and low RM and $\chi^2$ values.  Fig.\,\ref{rmfit}
shows the fits computed in such positions, indicated in
the inserted image. The data are well represented by a linear $\lambda^2$
relation. These results are in agreement with the interpretation that external
Faraday rotation is the dominant mechanism in the sources, although a much larger
$\lambda^2$ range, measured with a finer sampling, would be needed to confirm this 
hypothesis unambiguously.

Tab.~\ref{rmtab} reports, 
for the two sources (both separate and together), the  \rmm\, the
\srm\ and the maximum ($\arrowvert {\rm RM_{max}} \arrowvert$)
 absolute value  of the RM distribution.
These data were not corrected for the Galactic contribution, 
which is probably negligible. In fact, in galactic coordinates A2382
is located at lon=$38.91$\degrees~ and 
lat=$-46.93$\degrees~ and based on the average RM for extragalactic
sources published by Simard-Normandin et al. (1981),
the RM Galactic contribution in a region of about $10\degr $  	
centered on  A2382 is expected to be
about $-$5\,rad/m$^2$.

The RM results are consistent with the interpretation that the external Faraday screen is mostly due to the intracluster medium, indeed the source located in projection nearest to the cluster centre, i.e. PKS 2149-158C,  also has a higher RM.

The RM structures on small scales can be explained by the fact
that the cluster magnetic field fluctuates on scales
smaller than the size of the sources.
These results  suggest that it is necessary to consider  a cluster magnetic field 
that fluctuates over a wide range of spatial scales.

\begin{table}
\caption{Rotation measure values.}
\label{rmtab}
\centering 
\begin{tabular} {l c r c c}
\hline\hline
Source   & Distance & $<{\rm RM}>$ & $\sigma_{RM}$ & $\arrowvert {\rm RM_{max}}\arrowvert$ \\
         & (kpc) & (rad/m$^{2}$) & (rad/m$^{2}$) & (rad/m$^{2}$) \\
\hline 
PKS 2149-158  & 340 & 0.8    & 31 & 150  \\
PKS 2149-158C & 300 & $-$4.2 & 46 & 177 \\           
\hline
Both sources & - & $-$ 0.5 & 35 & 177 \\
\hline
\multicolumn{5}{l}{Col. 1: Source; Col. 2: Projected distance from the X-ray centroid;}\\
\multicolumn{5}{l}{Col. 3: Mean of the RM distribution; Col. 4: RMS of the }\\
\multicolumn{5}{l}{RM distribution; Col. 5: Maximum absolute value of the }\\
\multicolumn{5}{l}{RM distribution.}\\
\end{tabular}
\end{table}

\section{RM analysis: characterization of the intracluster magnetic field power spectrum}
\label{sec:RMan}
\begin{figure*}[!hb]
\centering
\includegraphics[width=16cm]{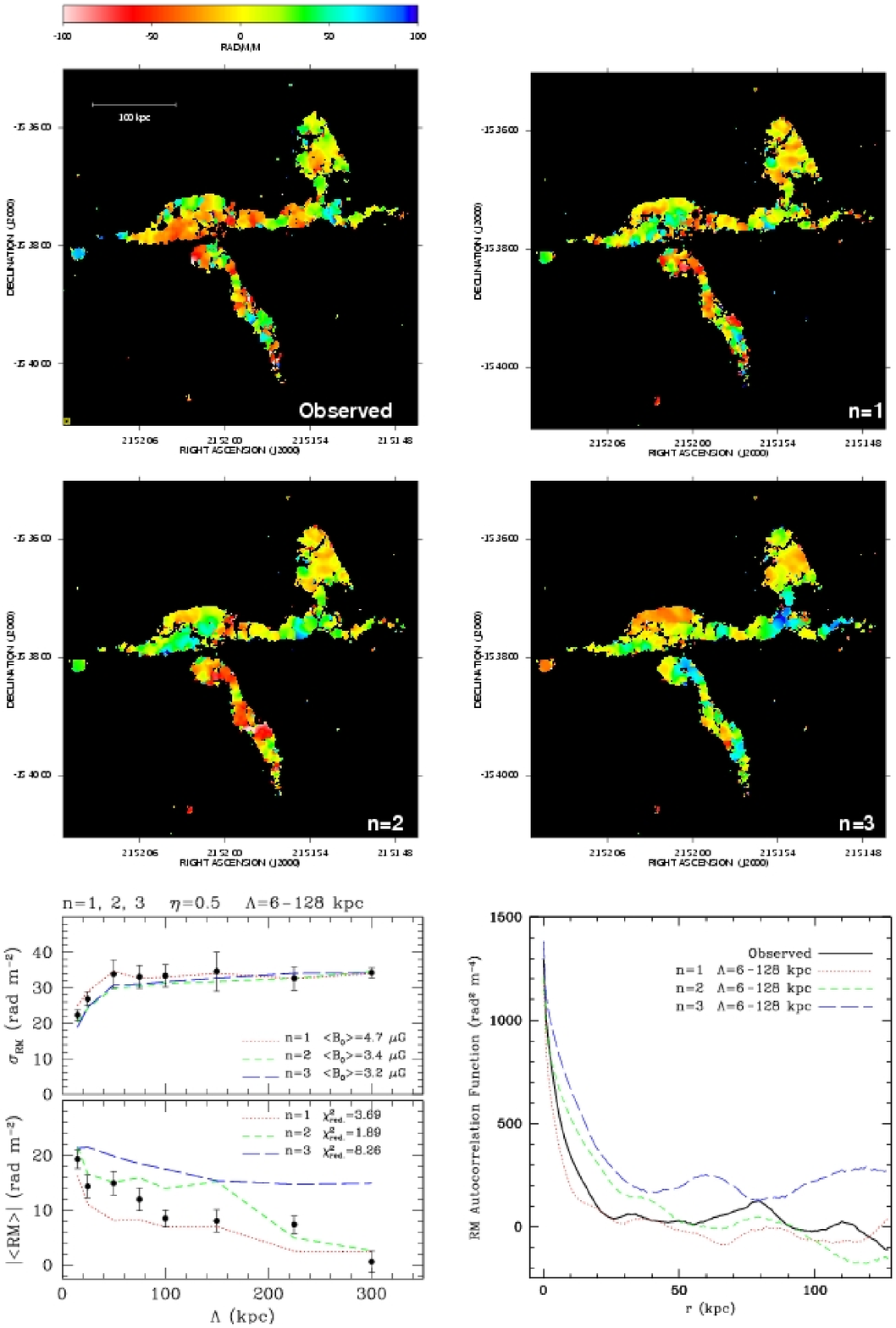}
\caption[]{Comparison of observed and simulated RM images of the sources PKS 2149-158 and PKS 2148-158C in A2382.
The observed RM image is shown in the top left panel.  The simulated images correspond to the value of $\Lambda_{max}=128$ kpc for
 the three values of spectral index $n=$1, 2, and 3. The model with $n=2$ kpc reproduces the observed 
RM statistics and auto-correlation function, which are shown in the bottom panels, best.}
\label{rmsim1}
\end{figure*}

The software package FARADAY (Murgia et al. 2004) permits the study of cluster magnetic fields 
by comparing the observed RM with simulated RM images obtained by considering 
three-dimensional multi-scale cluster magnetic field models.
In fact, given a three-dimensional magnetic field model and the density distribution of the intra-cluster gas,
FARADAY calculates the expected RM image by integrating Eq.\,\ref{equaz} numerically. In the specific case of A2382, 
the integration is performed from the cluster center up to three core radii ($\sim 1.1$ Mpc) along
the line-of-sight, i.e. both sources are supposed to lie in a plane which is perpendicular to line-of-sight and
intercepts the cluster centre. In the following we neglect the three-dimensional structure of the radio sources 
and assume that all the Faraday rotation occurs in the intracluster medium in between us and the sources.

The software has been adapted in such a way that it is now possible
to treat the simulations in the same way as the observations, i.e., using the same fit procedure that is used to derive 
the observed RM.
In particular, by using a source model of the intrinsic fractional polarization, polarization angle,  
and the observed I, U, and Q images, the task \textit{RM\_to\_UQ} produces, at each frequency and with 
the same noise as the data, the expected Q and U images corresponding to the simulated RM. Furthermore, 
we can take into account both the beam and bandwidth depolarization effects.
The synthetic U and Q images can be then processed by the task \textit{UQ\_to\_RM}, resulting in
final simulated RM images which are filtered with the same algorithm as the observations.
 In this way, the properties of the noise in the simulated RM images are very close to those of the
data. In particular, non-linear effects produced by the fit procedure, such as those due to the $n\pi$-ambiguities,
 are included in the simulations as well.

\subsection{The magnetic field model}
\label{subs:B}
We considered the power spectrum of the cluster magnetic field to be a power law of the type\footnote{Note that throughout this 
paper the power spectra are expressed as vectorial forms in $k$-space. 
The one-dimensional forms can be obtained by multiplying by $4\pi k^{2}$ and $2\pi k$
 respectively the three and two-dimensional power spectra. According to this notation the Kolmogorov 
spectral index is $n=11/3$.}
\begin{equation}
|B_k|^2\propto k^{-n}
\label{bpower}
\end{equation}
in a three-dimensional cubical box. The simulations begin in the Fourier space by extracting the amplitude of 
the magnetic field fluctuations from a Rayleigh distribution whose standard deviation
varies with the wave number according to Eq.\,\ref{bpower}. The phase of the magnetic field fluctuations
is random. We then performed a three-dimensional Fast Fourier Transform (FFT) inversion to produce the magnetic field in the real space domain. 
The field is Gaussian and isotropic, in the sense that there is no privileged direction in space for the magnetic field fluctuations.
We used a grid size of 1024$^3$ pixels in size in the wave number (Fourier) domain, which allows us to study the magnetic field fluctuations 
over a range of spatial scales more than two-order of magnitudes wide \footnote{Here we refer to 
the length $\Lambda$ as the magnetic field reversal scale. In this way, $\Lambda$ corresponds to a half-wavelength, i.e. $\Lambda= 0.5\cdot(2\pi/k)$.}.

In this set of simulations, the slope and the range of spatial scales of the magnetic field fluctuation is the same
 all over the cluster volume. However, the normalization of the power spectrum decreases with the distance from the
 cluster center. In particular, the average magnetic field strength varies according to:

\begin{equation}
\langle\mathbf B\rangle(r)=\langle B_{0}\rangle\left[\frac{n_e(r)}{n_0}\right]^{\eta}
\label{br}
\end{equation}

where $\langle B_{0}\rangle$ is the average magnetic field strength at the cluster centre while $n_{e}(r)$ is the thermal electron gas density 
 assumed to follow the double $\beta$-model profile described in Sect.\,\ref{subs:fitsx}.
 
The adopted magnetic field model has five free parameters (see Tab.\,\ref{param}): $B_{0}$, $n$, $\Lambda_{min}$, $\Lambda_{max}$, and $\eta$.

\begin{table}[]
\caption{Magnetic field model parameters.}
\label{simul}
\centering
\begin{tabular}{l l}
\hline
\hline                
 $B_{0}$              &    Average magnetic field strength at the cluster centre   \\ 
 $n$                  &    Power spectrum spectral index; $|B_k|^2\propto k^{-n}$ \\
 $\Lambda_{\rm min}$  &    Minimum scale of the magnetic field fluctuations \\
 $\Lambda_{\rm max}$  &    Outer scale of the magnetic field fluctuations \\
 $\eta$               &    Magnetic field radial profile slope; $\langle\mathbf B\rangle(r)=\langle B_{0}\rangle\left[\frac{n_e(r)}{n_0}\right]^{\eta}$  \\     
\hline
\label{param}
\end{tabular}\end{table}

By varying all these parameters we obtain synthetic RM images characterized by very different statistics and structure.
Our purpose is to find the combination of model parameters that gives the best representation of the observed distribution
of $\sigma_{RM}$ and $\langle RM \rangle$ across the sources and as well as their RM auto-correlation function.

Ideally, one would like to fit all the five free parameters simultaneously. However, in our case this is not very practical,
because of the computational burden caused by the FFT inversion. Therefore
we performed a series of simulations that search the best magnetic field power spectrum by varying at most one or two parameters at a time, 
while keeping the others fixed. We found that there are two main degeneracies between the model parameters. 
The first one is between $n$ and $\Lambda_{max}$: the higher is $n$ the lower is $\Lambda_{max}$. The second 
one is between $\eta$ and $B_{0}$: the higher is $\eta$ the higher is $B_{0}$. This means that  different
combinations of these parameters may yield an equally good fit to the data. 

In Sect.~\ref{subsub:first} we show the results obtained first by fixing $\Lambda_{min}$ and
$\Lambda_{max}$ while varying $n$. Then we give the results obtained  by fixing the spectral index
at the Kolmogorov value ($n$=11/3) while varying $\Lambda_{max}$.
In both cases we considered $\Lambda_{min}=6$\,kpc and $\eta$=0.5.
The choice for $\eta$ is motivated in Sect.~\ref{subsub:second}, where we also analyze 
how the $\eta$ parameter affects the magnetic field strength.
The choice for $\Lambda_{min}$ is supported by observations that reveal RM fluctuations on small scales.
However, in Sect.~\ref{subsub:third} we analyze the effect of the magnetic field minimum scale
on the polarization properties of the observed radio galaxies.

\subsection{RM statistics and auto-correlation function}\label{subsub:first}

In the following we compare the simulated and observed RM images. In order to 
assess if a given magnetic field power spectrum is able to reproduce the data, we considered
 two approaches: i) we analyzed the RM statistics ($\sigma_{RM}$ and $\langle RM \rangle$) calculated over areas
 of increasing size, and ii) we compared the RM auto-correlation functions.

In order to calculate the RM statistics, we covered the RM images with a regular grid of boxes of a given size. 
We then calculated a global average of all the $\sigma_{RM}$ and $\langle RM \rangle$ values found in each box. 
By changing the size of the boxes in the grid we obtained a trend of the average $\sigma_{RM}$ and $\langle RM \rangle$ 
as a function of the box size. We varied the size of the boxes from a minimum size of 15 kpc (49 boxes) up to a maximum size of 
300 kpc (1 box). 

The RM auto-correlation function is calculated as
\begin{equation}
A(r)=<RM(x,y)\cdot RM(x+dx, y+dy)>
\label{acf}
\end{equation}
where $r=\sqrt{dx^{2}+dy^{2}}$, while the average is taken over all the positions $(x,y)$ in the RM images, excluding 
blanked pixels. It is worth mentioning that $A(0)=\langle RM^{2}\rangle= \sigma_{RM}^{2}+\langle RM \rangle^{2}$.

In Fig.\,\ref{rmsim1} we show the results of a set of simulations obtained by fixing $\eta=0.5$, $\Lambda_{min}=6$ kpc, 
$\Lambda_{max}=128$ kpc, and by varying $B_{0}$ for three different values of $n=1, 2, 3$. The choice of $\eta$ and 
  $\Lambda_{min}$ is discussed in Sections \ref{subsub:second} and \ref{subsub:third}, respectively. The value of $\Lambda_{max}$ has been arbitrarily fixed at 128 kpc. 
This choice is motivated by the evidence of a zero \rmm in both radio galaxies, which indicates that the largest scales of 
the magnetic field fluctuations are smaller than the source size. In the top and middle panels of Fig.\,\ref{rmsim1} we present
the observed and the simulated RM images, using the same colour scale, cellsize and resolution. 
In the bottom left and right panels we show the RM statistics and the RM auto-correlation functions, respectively.

\begin{figure*}[h]
\centering
\includegraphics[width=16cm]{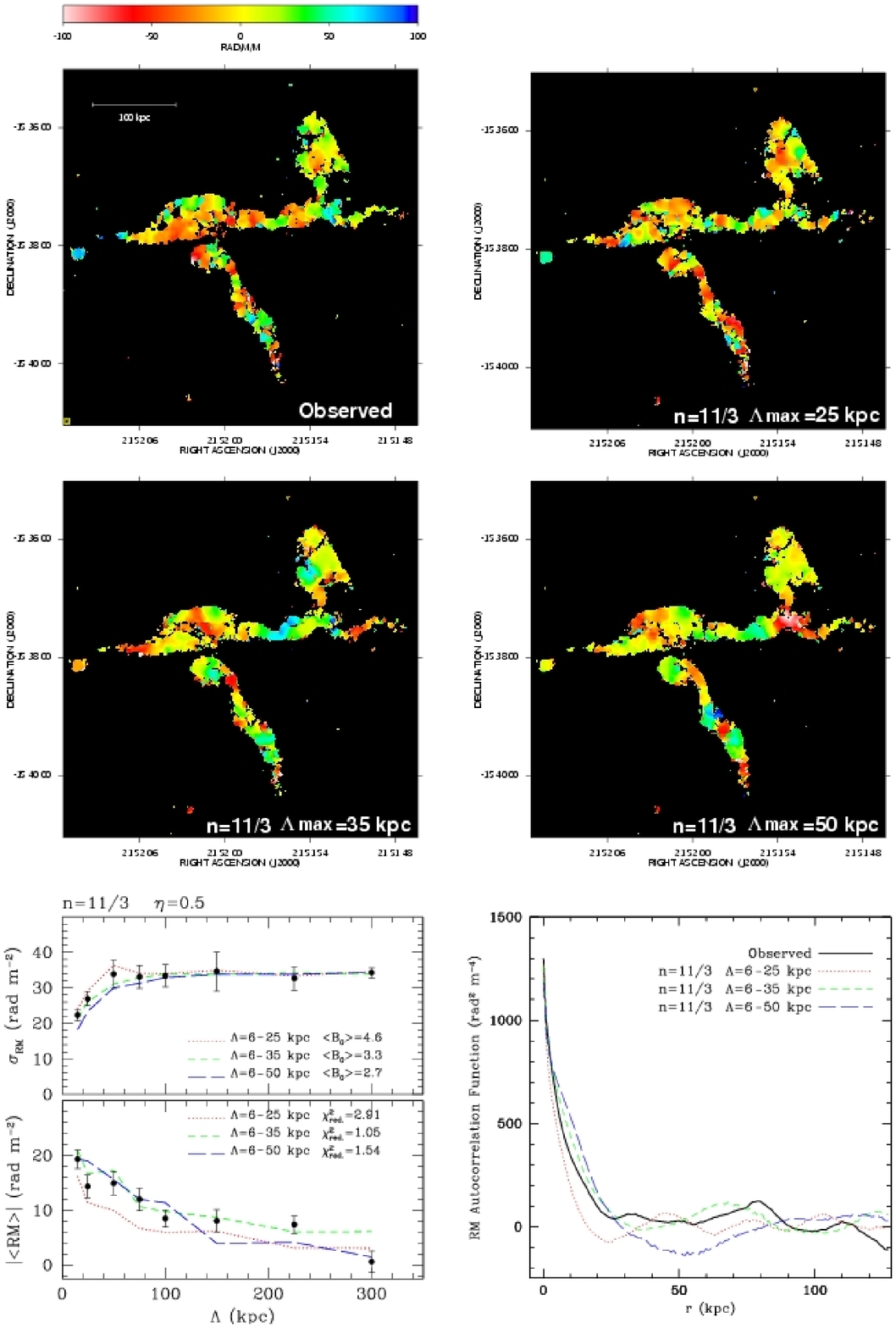}
\caption[]{Comparison of observed and simulated RM images of the sources PKS 2149-158 and PKS 2148-158C in A2382.
The observed RM image is shown in the top left panel. The simulated images correspond to the Kolmogorov index $n=11/3$ for
 the three values of  $\Lambda_{max}=$25, 35, and 50 kpc. The model with $\Lambda_{max}=35$ kpc reproduces the observed 
RM statistics and auto-correlation function, which are shown in the bottom panels, best.}
\label{rmsim2}
\end{figure*}

The global \srm calculated over sources is our most reliable statistical indicator, since
 it is based on a large number of independent measurements. Thus, in the fit procedure we first attempt to reproduce the
 \srm of the largest box in the statistics (the 300 kpc dot) by adjusting $\langle B_{0}\rangle$ for the three power spectra. We obtained a good 
fit of the \srm trend for all values of $n$ with a central magnetic 
field strength $\langle B_{0}\rangle$ in the range 3.2~--~4.7\,$\mu G$. However, it is clear that $n=3$, where most of the power of the
 RM fluctuations are concentrated on $\Lambda_{max}$, results in a \rmm level that is much higher than observed, while $n=1$, where the strongest
 RM fluctuations are on $\Lambda_{min}$, generates an \rmm lower than observed. The analysis of the \rmm trend  
 suggests that if $\Lambda_{max}=128$ kpc the power spectrum spectral index should be close to $n=2$. This is quantitatively confirmed
by the values of the reduced $\chi^{2}$ reported in bottom left panel of Fig.\,\ref{rmsim1}. Following Govoni et al. (2006), the $\chi^{2}$
 has been calculated according to

\begin{equation}
\chi^{2}=\sum \frac{(|\langle {\rm RM}_{\rm obs}\rangle |-|\langle {\rm RM}_{\rm sim}\rangle|)^{2}}{{\rm err}_
{|\langle {\rm RM}_{\rm obs}\rangle |}^{2}+{\rm err}_{|\langle {\rm RM}_{\rm sim}\rangle |}^{2}}~
\label{chi2}
\end{equation}

where the errors in the denominator take into account the statistical uncertainties in the data as well as in the simulations\footnote{For graphical reasons, 
the simulated trends in left bottom panels of Figs.\,\ref{rmsim1} and \,\ref{rmsim2} are represented as lines. However, the RM statistics of the
 simulated images, which has been calculated with the same set of boxes used for the data, is affected by the same uncertainty. Two effects 
contribute in determining the amplitude of the scatter. The first one is due to the error measurements while the second one is due to the statistical
 variance due to the random nature of the magnetic field in the Faraday screen. In our case, the latter is dominant over the former.}.

The same behaviour is seen in the RM auto-correlation functions shown in the bottom right panel of Fig.\,\ref{rmsim1}. A magnetic field 
power spectrum characterized by a spectral index $n=3$ and  $\Lambda_{max}=128$ kpc has too much power at all scales, compared to the data.
On the other hand, a magnetic field power spectrum characterized by a spectral index $n=1$ generates an RM image whose auto-correlation function
 lies below the observed one over most of the considered range of scales. The RM auto-correlation function corresponding to the intermediate
 case $n=2$ gives a better description of the data, confirming the result found on the RM statistics.

Even if the data can be quite well explained by a flat ($n=2$) and a broad ($\Lambda$=6~--~128 kpc)
magnetic field power spectrum, because of the degeneracy existing between $n$ and $\Lambda_{\rm max}$
the observed RM can be explained also by a narrower and steeper power spectrum. We produced 
a second set of simulations in which we kept the slope of the power spectrum fixed at the Kolmogorov
 value, $n=11/3$, and let $\langle B_{0}\rangle$ vary, for three different values of $\Lambda_{max}=25, 35, 50$ kpc.
The values of $\eta$ and $\Lambda_{min}$ are the same as in the previous set of simulations. 

In Fig.\,\ref{rmsim2} we compare the observed and simulated RM images corresponding to $n=11/3$.
We found that $\langle B_{0}\rangle$ falls in the range 2.7~--~ 4.6\,$\mu G$, a result which is very close to
 what was found previously. However in this case the degree of similarity between the simulated and the observed RM images remarkable.
As can be seen in the bottom panel of  Fig.\,\ref{rmsim2} the Kolmogorov models reproduce the data better than the wider and flatter 
power spectra considered above. In particular, the magnetic field power spectrum characterized by $n=11/3$ and $\Lambda_{max}=35$ kpc provides 
an excellent fit to the observed \rmm profile, yielding a reduced  $\chi^{2}$ close to unity. The Kolmogorov model with $\Lambda_{max}=25$ kpc 
does not have enough power on large scales to reproduce the observed \rmm levels. The model with $n=11/3$ and $\Lambda_{max}=50$ kpc provides 
a good fit to the \rmm statistics but fails to reproduce the observed \srm values on scales below 20 kpc.

This is further confirmed by the analysis of the RM auto-correlation functions shown in the right panel of Fig.\,\ref{rmsim2}.
The RM auto-correlation function of the Kolmogorov power spectrum with $\Lambda_{max}=35$ kpc is very similar to the observed one.
The Kolmogorov power spectrum with $\Lambda_{max}=25$\,kpc has too much power on small scales and its auto-correlation function cuts off faster than 
the observed one. 
The model with $\Lambda_{max}=50$ kpc has too much power below 20 kpc and cuts off too late in terms
 of spatial scales, thus failing to reproduce the observed RM auto-correlation function.

To summarize, the analysis of the RM statistics and auto-correlation functions reveals that the best fit to the data is
obtained by a Kolmogorov power spectrum with $\Lambda_{max}=35$ kpc and $\langle B_{0}\rangle=3.3\,\mu G$.

\subsection{The magnetic field strength radial profile}\label{subsub:second}
The amount of RM depends on the integral of the product of the field intensity and the 
electron density along the line-of-sight (see Eq.\,\ref{equaz}). This dependency results in
 a degeneracy between the $\eta$ and $\langle B_0 \rangle$ parameters. Although 
$\eta$ does not dramatically affect the estimate of the magnetic field power spectrum spectral
index, it can however strongly affect the estimate of the magnetic field strength.
 In particular, the steeper 
the magnetic field radial trend, i.e. the higher is $\eta$, the higher should be $\langle B_0 \rangle$
in order to reproduce the observed RM levels.

Here we justify the choice $\eta=0.5$ adopted in Sect~\ref{subsub:first}, and we critically discuss 
the estimate of the magnetic field central strength as a function of the $\eta$ parameter.  
 
It is possible to obtain a constraint on the index $\eta$ and therefore  allowing measurement of the magnetic
field distribution in A2382, by comparing the observed and simulated \srm profile as function
of distance from the cluster centre.

\begin{figure*}[t]
\centering
\includegraphics[width=18cm]{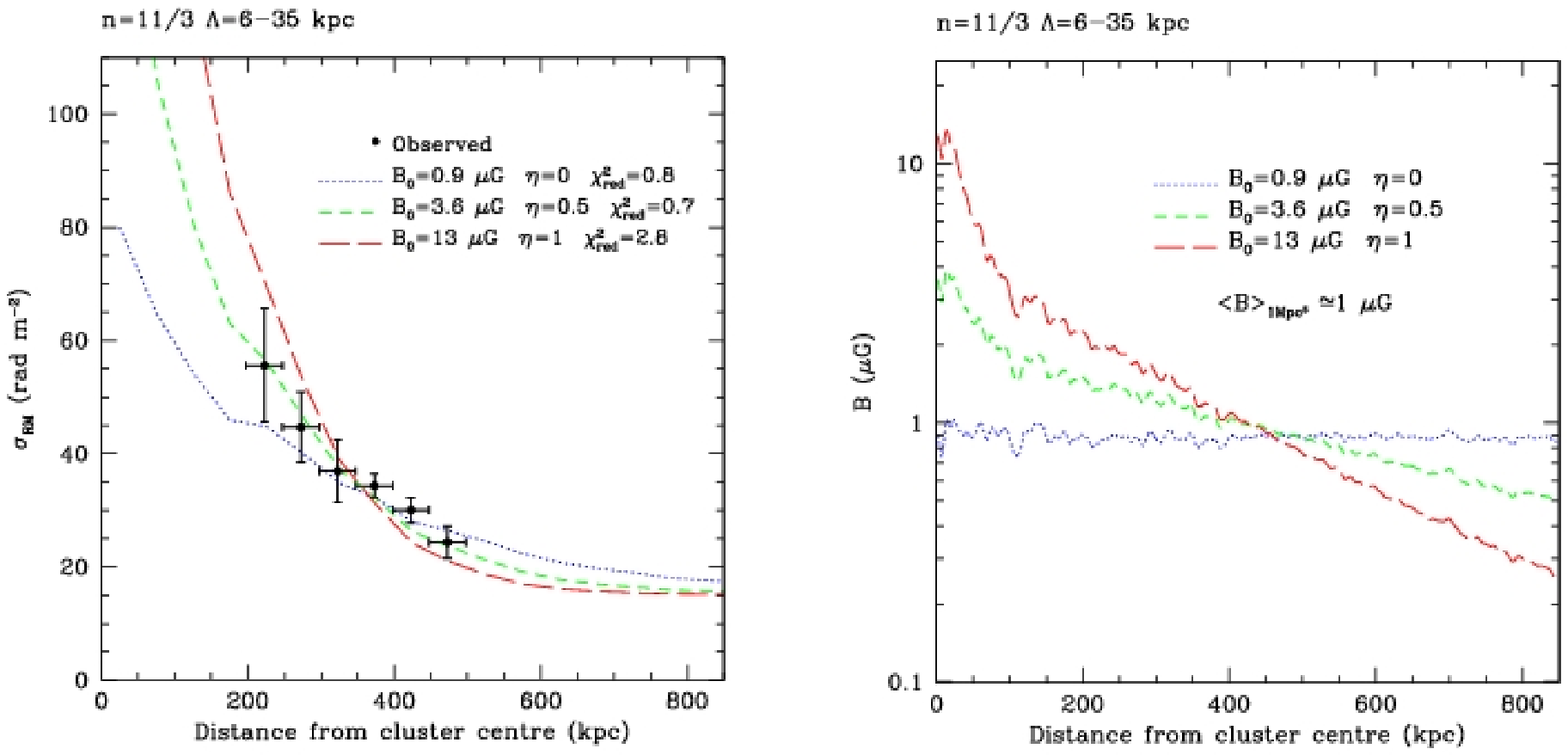}
\caption[]{Left panel: Radially averaged $\sigma_{RM}$ profile as a function of the
 projected distance from the cluster center. Solid lines represent the trends obtained from the simulation 
for a magnetic field configuration with $n=11/3$ and $\Lambda_{max}=35$ kpc for different values of the $\eta$ parameter. Right panel: magnetic field 
strength profile for the three values of the $\eta$ parameter. The average magnetic field strength
 over the central 1\,Mpc$^3$ is about 1\,$\mu G$ for all the three models.}
\label{eta}
\end{figure*}

In Fig.\,\ref{eta} we show the observed \srm profile compared with the simulations
obtained for different values of the $\eta$ parameter, for the model which best
reproduces the observed RM image statistics and auto-correlation function, i.e. $n=11/3$, $\Lambda_{min}=6$ 
kpc, and $\Lambda_{max}=35$ kpc. The radial profiles shown in  Fig.\,\ref{eta} have
 been traced by calculating \srm in 50 kpc wide concentric annuli centred on the 
cluster X-ray centroid. The simulated radial profiles were extended to both smaller and larger distances
 from the centre compared to the range of distances covered by PKS 2149-158 and PKS 2148-158C. 
For this reason, and different from the procedure described in Sect.\,\ref{subsub:first}, the simulated RM images
were not filtered the same way as the observed images. Instead, we added in quadrature to the simulated \srm a constant RM noise of 
 15 rad m$^{-2}$. This value is the average noise introduced by the fit procedure in the case of the considered
 magnetic field model. 

The analysis of the radially average \srm profiles presented in Fig.\,\ref{eta} shows that the best fit of
the data is obtained with $\eta$=0.5 and $\langle B_0 \rangle=3.6\, \mu G$. This value of $\eta$ corresponds to a magnetic field whose energy density 
decreases from the cluster center as the square root of the gas electron density. The value of the central magnetic field strength obtained 
through the fit of the radially averaged \srm profile is indeed in very good agreement with that obtained by the fit of the RM statistics, 
$\langle B_0 \rangle=3.3\, \mu G$. A constant magnetic field, represented by the $\eta=0$ profile, with a strength of
 $\langle B_0 \rangle=0.9\, \mu G$ also provides a good fit of the data, although the $\chi^{2}$ is slightly worse than the case corresponding 
 to $\eta$=0.5. The steepest magnetic field radial profile considered here, $\eta=1$, provides the highest central magnetic field strength 
 $\langle B_0 \rangle=13\, \mu G$ but also the worst fit of the data.
The formal uncertainty in the central magnetic field as provided by the fit procedure, {\it i.e. excluding all the systematic effects and 
 by keeping fixed all the other four parameters}, is of the order of
 $ \langle B_0 \rangle=3.6 \pm 0.5\,\mu G$. However, Fig.\,\ref{eta} shows that if we allow both $\eta$ and $\langle B_0\rangle$ to vary the 
$1-\sigma$ confidence region around the best fit parameters is certainly larger and a safer estimate is to consider the central magnetic field
 strength in the range  $1< \langle B_0 \rangle < 13 \, \mu G$.

It should be noted that the above conclusions are based on the assumption that  PKS 2149-158 and PKS 2148-158C lie at the same distance
 along the line-of-sight. On the other hand, even if different values of the $\eta$ parameter
 lead to quite different values for the central magnetic field strength, the average magnetic field strength over a larger cluster volume
%Hans: non capisco "for the three index ..."
 is nearly the same for the three indices $\eta$. In fact, the average magnetic field strength over the central 1\,Mpc$^3$ 
 is almost the same for all the three models:  $\langle B \rangle_{\rm 1\,Mpc^3}\simeq 1\, \mu G$.

\subsection{The depolarization constraint on the magnetic field minimum scale}\label{subsub:third}

\begin{figure*}
\centering
\includegraphics[width=16cm]{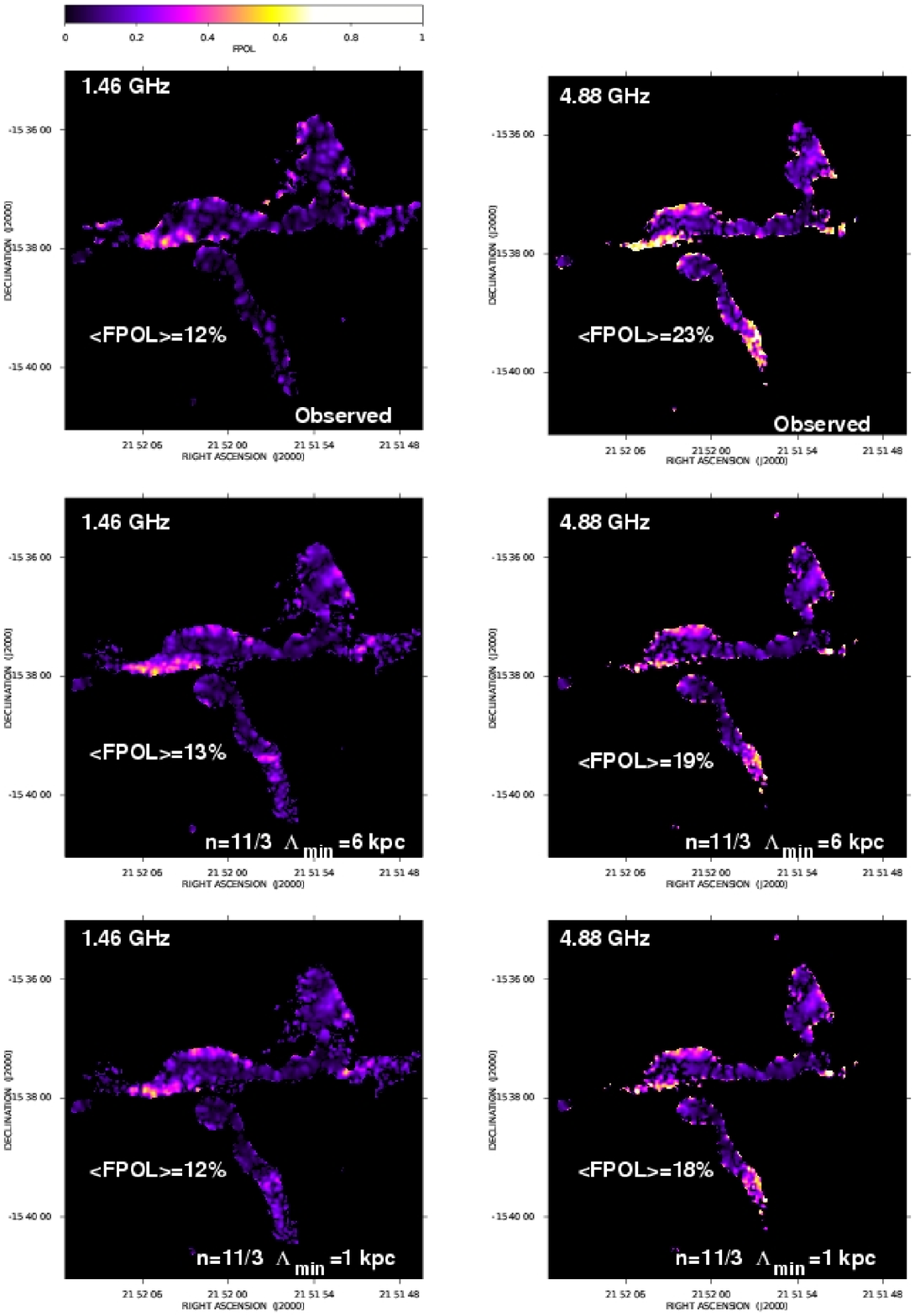}
\caption[]{Observed and simulated depolarization for the Kolmogorov power spectrum. The fractional polarization averages comprise both the sources.}
\label{dpnk}
\end{figure*}

In all the models considered so far we fixed the value of the minimum scale of magnetic field fluctuations to  $\Lambda_{min}=6$ kpc. 
This choice was motivated by the fact that this scale matches the linear resolution of the radio images, i.e. $6.2\times6.2$ kpc. 
Here we evaluate the effects on the simulated RM images by considering a $\Lambda_{min}$ that is smaller than the beam. The size of the
computational box allow us to push the value of $\Lambda_{min}$ down to 1 kpc.
The most direct consequence arising from a magnetic field power spectrum which fluctuates on scale smaller than the linear resolution
of the  observations is the beam depolarization effects. 
With the same FARADAY tool used to filter the simulations as well as the observations, it is possible to translate an RM image into 
U and Q images at any given frequency and resolution. This allows us to estimate the amount of beam depolarization 
resulting from different value of $\Lambda_{min}$. In Fig.\,\ref{dpnk} we show the comparison of the observed and simulated
fractional polarization in the case of the best fit magnetic field model with $n=11/3$, $\eta=0.5$, and $\Lambda_{max}=35$ kpc.
Middle and bottom panels of Fig.\,\ref{dpnk} show the simulated fractional polarization for $\Lambda_{min}=6$ and 1 kpc, respectively.
The amount of depolarization between 4.88 and 1.46 GHz is not only in good agreement with the data, but also
 almost the same for the two scales. In the case of  the Kolmogorov spectrum the beam depolarization is therefore very weak.
This can be explained by the fact that, for such a steep spectrum model, most of the magnetic energy density resides in the large scales fluctuations.
Thus, the simulated RM images are almost insensitive to changes on $\Lambda_{min}$. However, for the same reason we cannot put a lower limit on it.
It should be considered that lowering $\Lambda_{min}$ results in an higher magnetic field strength. 
In the case of the Kolmogorov spectrum with $n=11/3$, $\eta=0.5$, and $\Lambda_{max}=35$ kpc, by lowering $\Lambda_{min}$ from 6 to 1 kpc 
requires an increase in the magnetic field strength from  $\langle B_0 \rangle=3$ to $ 5\,\mu G$ on order to explain the observed RM values. This is due 
 to the fact that the magnetic field auto-correlation length, $\Lambda_{B_{z}}$, is also smaller and since the RM scales as 
$\langle RM^2\rangle \propto \langle B_0 \rangle\cdot \Lambda_{B_{z}}^{1/2}$ (see Eq.\,15 in Murgia et al. 2004) we need to increase 
$\langle B_0 \rangle$ if $\Lambda_{B_{z}}$ is lowered.

\begin{figure*}
\centering
\includegraphics[width=16cm]{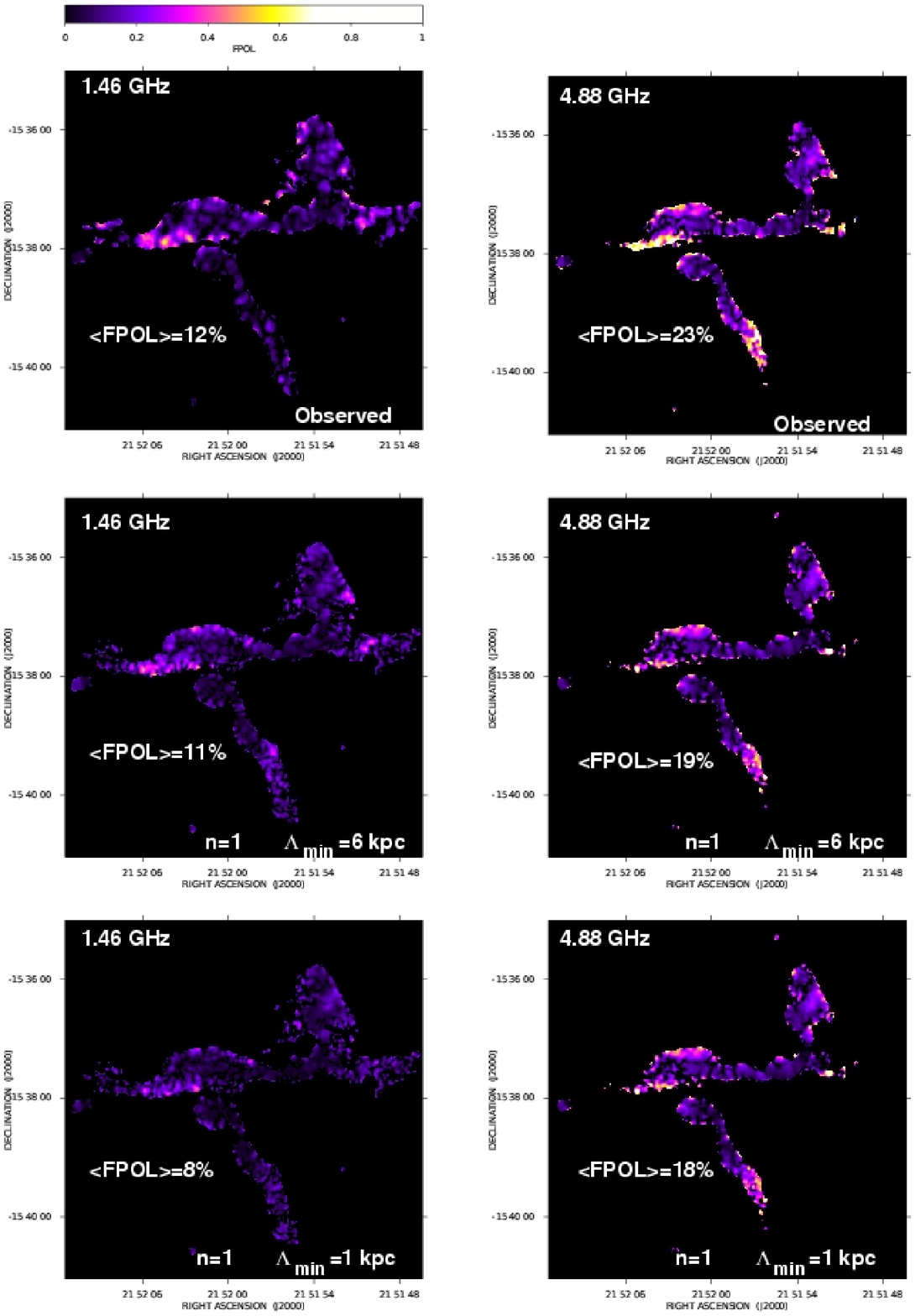}
\caption[]{Observed and simulated depolarization for the shallow power spectrum with $n=1$. The fractional polarization averages comprise both the sources.}
\label{dpn1}
\end{figure*}

The same considerations apply to the model with $n=3$ and $n=2$. The situation is different for $n=1$. In this case most 
of the power of the RM fluctuation is concentrated in the small scales. Therefore, lowering $\Lambda_{min}$ leads to a significant beam
 depolarization. This is illustrated in Fig.\,\ref{dpn1}, where we report the simulated fractional polarization in the case of the 
shallow power spectrum with  $n=1$, $\eta=0.5$, and $\Lambda_{max}=128$ kpc. The expected fractional polarization at 1.46 GHz decreases from 
 about 11\% down to 8\% when lowering $\Lambda_{min}$ from 6 to 1 kpc. Thus for a flat power spectrum with $n\le 2$, $\Lambda_{min}$ should not be smaller than the linear resolution of the  observations in order to prevent the depolarization of the signal at low frequencies.

\section{Summary and conclusions}
\label{sec:sum}
In this work we studied the strength and structure of the magnetic field in the cluster of galaxies A2382.
Following a numerical approach we investigated the relationship between magnetic field and Faraday rotation effects in this cluster.
For this purpose, we presented Very Large Array observations at 20\,cm and 6\,cm of two polarized radio sources embedded in  A2382, and we
obtained detailed rotation measure images for both of them. We analyzed the X-ray emission of A2382 observed by the ROSAT satellite and 
derived the radial profile of the electron gas density. The observed X-ray surface brightness profile 
cannot be described by a simple $\beta$-model due to the presence of a core of strong emission in the inner 100\,kpc of the cluster.
A double $\beta$-model provides a better fit of the X-ray surface brightness profile. There is indeed the possibility that A2382 is
a cooling core cluster, although new spectroscopic X-ray observations are required in order to confirm the hypothesis.

We simulated random three-dimensional magnetic field models with different power
spectra and produced synthetic RM images. We filtered the synthetic RM images with the same fit procedure used to derive the observed 
RM images in order to ensure a proper comparison of the simulations with the data. By comparing our simulations with the observed 
polarization properties of the radio sources, we determined the strength and the power spectrum of intra-cluster magnetic field fluctuations
which best reproduce the observation. 

By assuming that PKS 2149-158 and PKS 2148-158C lie at the same distance along 
the line-of-sight and neglecting their three-dimensional structure, we conclude that the data are consistent with a 
power law magnetic field power spectrum with the Kolmogorov index $n=11/3$ while the largest scales of the magnetic field fluctuations 
are of the order of 35\,kpc. The average magnetic field strength at the cluster center is 
about 3\,$\mu$G and decreases in the external region as the square root of the electron gas density. The average magnetic field strength 
over the central 1\,Mpc$^{3}$ is about 1\,$\mu$G.

\begin{acknowledgements}
We would like to express our gratitude to Gabriele Giovannini for the critical reading of the manuscript, and 
to Tiziana Venturi and Marco Bondi for many helpful discussions.
D.\,G. wish to thank the staff of the Astronomical Observatory of Cagliari for their
hospitality and support during the development of this work. This research has been partially funded by 
the Grant ASI-INAF I/088/06/0 - High Energy Astrophysics. We made use of the NASA/IPAC Extragalactic 
Database (NED) which is operated by the Jet Propulsion Laboratory, California Institute of Technology, under
contract with the National Aeronautics and Space Administration.

\end{acknowledgements}

\end{document}